\documentclass[aps,twocolumn,amsmath,amssymb,nofootinbib,superscriptaddress,showpacs,floatfix,prb]{revtex4}

\usepackage{latexsym}
\usepackage{graphicx}
\usepackage{times,psfrag,subfigure}
\usepackage{amsmath}
\usepackage{dcolumn}
\usepackage{bm, bbm} 
\usepackage{color}
\usepackage{soul}
\usepackage{latexsym,amsmath,amssymb,bm,euscript}
\def\re#1{(\ref{#1})}

\newcommand{\beq}{\begin{equation}}
\newcommand{\eeq}{\end{equation}}
\newcommand{\beqarray}{\begin{eqnarray}}
\newcommand{\eeqarray}{\end{eqnarray}}

\begin{document}

\title{Resonant generation of coherent phonons in a superconductor by ultrafast optical pump pulses}

\date{\today}

\author{Andreas P. Schnyder}
\email{a.schnyder@fkf.mpg.de}
\affiliation{Max-Planck-Institut f\"ur Festk\"orperforschung, Heisenbergstrasse 1, D-70569 Stuttgart, Germany}

\author{Dirk Manske}
\affiliation{Max-Planck-Institut f\"ur Festk\"orperforschung, Heisenbergstrasse 1, D-70569 Stuttgart, Germany}

\author{Adolfo Avella}
\email{avella@physics.unisa.it}
\affiliation{Max-Planck-Institut f\"ur Festk\"orperforschung, Heisenbergstrasse 1, D-70569 Stuttgart, Germany}
\affiliation{Dipartimento di Fisica ``E.R. Caianiello'' - Unit\`{a} CNISM di Salerno, Universit\`{a} degli Studi di Salerno, I-84084 Fisciano (SA), Italy}
\affiliation{CNR-SPIN, UoS di Salerno, I-84084 Fisciano (SA), Italy}
 
\begin{abstract}
We study the generation of coherent phonons in a superconductor by ultrafast optical pump pulses.
The nonequilibrium dynamics of the coupled Bogoliubov quasiparticle-phonon system after excitation with the pump pulse is analyzed by means of the density-matrix formalism with the phonons treated at a full quantum kinetic level. For ultrashort excitation pulses, the superconductor exhibits a nonadiabatic behavior in which the 
superconducting order parameter oscillates. We find that in this nonadiabatic regime the generation of coherent phonons is resonantly enhanced when the frequency of the order-parameter oscillation is tuned to the phonon energy, a condition that can be achieved in experiments by varying the integrated pump pulse intensity.
\end{abstract}

\date{\today}


\maketitle

\section{Introduction}

The generation of coherent phonons by ultrashort optical pulses with duration much shorter than the phonon vibrational period has been extensively studied in various materials, such as bulk semiconductors, \cite{choKurzPRL90,zeigerPRB92,scholzPRB93,kuznetsov94,schilp94,kuznetsov95,garretPRL96,merlinReview97} semiconductor quantum wells\cite{dekorsy96,yeePRL2001,kojimaPRB04,papenkortPRB2010,papenkortIOP2010,papenkortPSS2011} and superlattices,\cite{dekorsyPRL00,ghoshPRL00} as well as high-temperature superconductors.\cite{agostinelli91,mazin94,misochko01,marsi09,rettig10} For semiconducting systems, several distinct coherent phonon generation mechanisms have been discussed.\cite{zeigerPRB92,kuznetsov94,schilp94,kuznetsov95,garretPRL96} For example, in the displacive mechanism, the optical pulse creates a finite photocarrier distribution almost instantaneously on the time scale of the phonon subsystem.\cite{zeigerPRB92,kuznetsov94} This results in an abrupt change of the equilibrium positions of the lattice ions, and hence gives rise to coherent oscillations of the atoms around the new potential minima. In the impulsive mechanism, an effective direct coupling of the laser field to the lattice ions is assumed,\cite{garretPRL96,liuMerlin95} leading to a brief and intense force acting on the atoms. The detailed dynamics of the electronic subsystem on time scales longer than the optical pulse is, in general, irrelevant for the description of coherent phonon creation in semiconductors. An exception to this rule occurs when the electronic subsystem oscillates with a period on the time scale of the phonon vibrations, in which case the coherent phonon generation is resonantly enhanced . This has been observed both in semiconductor quantum wells\cite{kojimaPRB04,papenkortPRB2010,papenkortIOP2010,papenkortPSS2011} and in superlattices.\cite{dekorsyPRL00,ghoshPRL00} 

In this paper, we investigate the generation of coherent phonons in a nonequilibrium superconductor. Specifically, we study the optical excitation of Bogoliubov quasiparticle states above the superconducting ground state on time scales shorter than the phonon vibrational period $\tau_{\textrm{ph}}$. We find that, akin to the displacive mechanism in semiconductors, a sudden change in the Bogoliubov quasiparticle distribution functions generates coherent phonon oscillations. 
This mechanism of phonon creation is relevant to pump-probe experiments on superconductors with a pump photon energy of the same order but slightly larger than twice the superconducting gap amplitude $|\Delta|$ and a laser pulse duration $\tau_{\textrm{p}}$ shorter than both the phonon period $\tau_{\textrm{ph}}$ and the dynamical time scale of the superconducting order parameter $\tau_{\Delta} \sim  h / (2 |\Delta|$).
It has recently been shown that, whenever the pump pulse duration $\tau_{\textrm{p}}$ is much shorter than $\tau_{\Delta}$, oscillations are  created in the quasiparticle 
occupations with frequency of the order of $\sim 2 \pi / \tau_{\Delta} \sim 2 |\Delta| /  \hbar$.\cite{barankovPRL04,amin2004,Yuzbashyan05,YuzbashyanJPA05,Yuzbashyan06,papenkort07,papenkort08,papenkort09,papenkortConf09}  For $\tau_{\Delta} \ll \tau_{\textrm{ph}}$, 
these oscillations average out on the time scale of the phonons and are therefore unimportant for the creation of coherent phonons. When  
$\tau_{\Delta}$ is close to $\tau_{\textrm{ph}}$, on the other hand, the generation of coherent phonons is resonantly enhanced. 
Remarkably, provided that $\omega_{\textrm{ph}} \lesssim 2 |\Delta| / \hbar$, the Bogoliubov quasiparticle oscillations can be brought into exact resonance with the phonon frequency $\omega_{\textrm{ph}}$ by adjusting the integrated pump pulse 
intensity (see Figs.~\ref{fig2} and \ref{fig3}).

In the following, we theoretically investigate this resonant coherent phonon generation mechanism by employing a microscopic model of an $s$-wave superconductor coupled to an optical phonon mode with frequency $\omega_{\textrm{ph}}$. We study the pulse-induced dynamics of this model system at times shorter than the quasiparticle energy relaxation time $\tau_{\epsilon}$, a regime which can be fully described within mean-field BCS theory.\cite{YuzbashyanJPA05,volkov74} Different orderings of the involved time scales are studied with a particular emphasis on the case where both the phonon and the quasiparticle subsystems evolve in a nonadiabatic fashion, i.e. where $\tau_{\textrm{p}} \ll \tau_{\textrm{ph}} , \tau_{\Delta} \ll \tau_{\epsilon}$. In this nonadiabatic regime, traditional approaches for computing nonequilibrium dynamics, such as the time-dependent Ginzburg-Landau theory or the Boltzmann kinetic equation, are not applicable, since the full dynamics of both the normal and the anomalous quasiparticle densities, as well as that of the coherent-phonon amplitudes needs to be accounted for. Therefore, we resort to the density-matrix formalism \cite{rossiKuhnRMP02,herbstKuhnPRB03} to numerically compute the coherent response of the model system after excitation by a short pump pulse. Based on this approach, we analyze in detail the generation of coherent phonons and calculate lattice displacements both for resonant and off-resonant conditions.
The analysis presented in this paper is complementary to the one of Ref.~\onlinecite{unterhin08}, which employs Boltzmann-type kinetic equations to study the adiabatic dynamics of Bogoliubov quasiparticles coupled to \emph{incoherent} phonons.

\section{Microscopic model}

The microscopic model we consider is a single-band BCS $s$-wave superconductor coupled to an external electromagnetic field and to a single branch of optical phonons $H = H_{\textrm{sc}} + H_{\textrm{em}} +H_{\textrm{ph} } + H_{\textrm{e-ph}}$. Within mean-field theory, the BCS superconductor is given by the following Hamiltonian
\begin{subequations} \label{model1}
\begin{eqnarray}
\label{BCSham}
H_{\textrm{sc}}
=
\sum_{{\bf k} , \sigma}
\varepsilon_{{\bf k}} c^{\dag}_{{\bf k}, \sigma} c^{\ }_{{\bf k}, \sigma}
+
 \sum_{{\bf k} \in \mathcal{W} } \left[
\Delta c^{\dag}_{{\bf k} \uparrow} c^{\dag}_{-{ \bf k} \downarrow}
+
\Delta^{\ast} c^{\ }_{-{ \bf k} \downarrow} c^{\ }_{{\bf k} \uparrow} 
\right] ,
\nonumber\\
\end{eqnarray}
where $c_{{\bf k},\sigma}$ represents the electron annihilation operator with spin $\sigma$ and momentum ${\bf k}$, $\varepsilon_{{\bf k}} = \hbar^2 {\bf k}^2 / (2 m) - E_{\textrm{F}}$, $m$ is the effective electron mass, and $E_{\textrm{F}}$ denotes the Fermi energy. The second sum in Eq.~\re{BCSham} is over the set $\mathcal{W}$ of momentum vectors with $| \varepsilon_{{\bf k}} | \leq \hbar \omega_{\textrm{c}}$, $\omega_{\textrm{c}}$ being the cut-off frequency. 
The superconducting order parameter $\Delta$ is assumed to have $s$-wave symmetry with $\Delta=W_0 \sum_{{\bf k} \in \mathcal{W} } \left\langle c_{- { \bf k} \downarrow} c_{+ { \bf k} \uparrow} \right\rangle$. Here, $W_0$ is an attractive momentum-independent interaction constant.

The superconducting system \eqref{BCSham} is perturbed by a Gaussian pump pulse, which creates finite nonequilibrium quasiparticle distributions. In the Coulomb gauge, the optical pump pulse is described by a transverse vector potential
\begin{eqnarray}
{\bf A}_{\bf q} ( t)
=
{\bf A}_0
e^{
- \left( \frac{ 2 \sqrt{ \ln 2} t }{ \tau_{\textrm{p}} } \right)^2 
}
\left(
\delta_{{\bf q}, {\bf q}_{\textrm{p}} } e^{- i \omega_{\textrm{p}} t}
+ 
\delta_{{\bf q}, - {\bf q}_{\textrm{p}} } e^{ + i \omega_{\textrm{p}} t }
\right) , 
\end{eqnarray}
with full width at half maximum (FWHM) $\tau_{\textrm{p}}$, amplitude ${\bf A}_{0}$, photon frequency $\omega_{\textrm{p}}$, and photon wave vector ${\bf q}_{\textrm{p}}$. The coupling
of the vector potential ${\bf A}_{{\bf q}}$ to the superconductor \eqref{BCSham} is given by
$H_{\textrm{em}} = H^{(1)}_{\textrm{em}} +H^{(2)}_{\textrm{em}}$, where
\begin{eqnarray}
\label{definition H em}
H^{(1)}_{\textrm{em}}
&=&
\frac{e \hbar }{2m} 
\sum_{{\bf k}, {\bf q}, \sigma}
( 2 {\bf k} + {\bf q} ) \cdot {\bf A}_{{\bf q}} (t) \; 
c^{\dag}_{{\bf k}+ {\bf q}, \sigma} c^{\ }_{{\bf k}, \sigma} ,
\\
H^{(2)}_{\textrm{em}}
&=&
\frac{e^2 }{2m} 
\sum_{{\bf k}, {\bf q}, \sigma}
\left[
\sum_{{\bf q}' } {\bf A}_{{\bf q}-{\bf q}' } (t) \cdot
{\bf A}_{{\bf q}' } (t) 
\right] c^{\dag}_{{\bf k}+ {\bf q}, \sigma} c^{\ }_{{\bf k}, \sigma} .
\nonumber
\end{eqnarray}

We add to this Hamiltonian a noninteracting phonon system and a coupling between the phononic and electronic degrees of freedom. The free-phonon Hamiltonian $H_{\textrm{ph}}$ 
is described by
$
H_{\textrm{ph} }
=
\sum_{{\bf p}} \hbar \omega_{\textrm{ph}} \left( b^{\dag}_{{\bf p}} b^{\ }_{{\bf p}} + \frac{1}{2} \right)
$,
where $b_{{\bf p}}$ is the annihilation operator of a phonon with wave vector ${\bf p}$ and constant frequency $\omega_{\textrm{ph}}$. For the sake of simplicity, we restrict ourselves to a single branch of phonons. A generalization to several phonon modes is straightforward. The superconductor is coupled to the phononic system via the interaction  
\begin{eqnarray} \label{ePhInt}
H_{\textrm{e-ph} }
&=&
g_{\textrm{ph} }
\sum_{{\bf p}, {\bf k}, \sigma }
( b^{\dag}_{-{\bf p}} + b^{\ }_{{\bf p}} ) c^{\dag}_{{\bf k}+{\bf p}, \sigma} c^{\ }_{{\bf k}, \sigma} ,
\end{eqnarray}
\end{subequations}
where $g_{\textrm{ph}}$ denotes the electron-phonon coupling constant.
In the following we assume that the electron-phonon coupling strength is much smaller 
than the superconducting energy scales,\cite{footnoteGph} such that the influence of the phonon subsystem on the
superconductor becomes negligibly small.

\vspace{0.9cm}

\section{Density-matrix formalism}
\label{densityM}

Physical observables, such as the order parameter amplitude $|\Delta(t)|$ and the lattice displacement $U({\bf r}, t)$, can all be expressed in terms of the Bogoliubov quasiparticle densities
and the mean phonon amplitudes. Hence, we derive equations of motion for these quantities using the framework of the density-matrix formalism. To this end, it is advantageous to perform a canonical Bogoliubov transformation of the fermionic operators, with $\alpha^{\ }_{{\bf k}} = u_{{\bf k}} c^{\ }_{{\bf k} \uparrow} + v_{{ \bf k}} c^{\dag}_{-{\bf k} \downarrow}$ and $\beta^{\dag}_{ {\bf k}} = u_{{\bf k}} c^{\dag}_{- {\bf k} \downarrow} - v_{ {\bf k}} c^{\ }_{ {\bf k} \uparrow}$, where the coefficients $u_{ { \bf k}}$ and $v_{ { \bf k}}$ are time-independent and chosen such that the BCS part of the Hamiltonian, $H_{\textrm{sc}}$, in the initial state, i.e., at $t = t_i$, takes diagonal form (see Appendix~\ref{appendixA}).
Due to the interaction term $H_{\textrm{e-ph}}$, Eq.~\eqref{ePhInt}, the equations of motion for the single-particle density matrices are not closed, but give rise to an infinite hierarchy of equations of higher-order density matrices. For the purpose of studying the generation of coherent phonons, it suffices to break this hierarchy at first order, which amounts to neglecting all correlations among quasiparticles and phonons. Thus, phonon-assisted quantities, such as $\langle \alpha^{\dag}_{{\bf k} } \alpha^{\phantom{\dag} }_{{\bf k}'} b^{\phantom{\dag}}_{{\bf p}} \rangle$, are factorized according to
$
\langle \alpha^{\dag}_{{\bf k} } \alpha^{\phantom{\dag}}_{{\bf k}'} b^{\phantom{\dag}}_{{\bf p}} \rangle
\simeq
\langle \alpha^{\dag}_{{\bf k} } \alpha^{\phantom{\dag}}_{{\bf k}'} \rangle \langle b^{\phantom{\dag}}_{{\bf p}} \rangle 
$. 
A nonvanishing $\langle b^{\phantom{\dag}}_{{\bf p}} \rangle$ corresponds to a finite displacement of the lattice ions. That is, the lattice displacement $U( {\bf r},t)$ is connected to the coherent-phonon amplitude $D_{\bf p} (t) = \langle b^{\phantom{\dag}}_{\bf p} \rangle + \langle b^{\dag}_{ -{\bf p} }\rangle$ via
\begin{eqnarray}
U ( {\bf r}, t )
=
\sqrt{ \frac{ \hbar }{ 2 M \omega_{\textrm{ph} } V } }
\sum_{{\bf p}}
D_{\bf p} (t) 
e^{+ i {\bf p} \cdot {\bf r} }, 
\end{eqnarray}
where $M$ is the reduced mass of the lattice ions and $V$ the systems volume.

At first order in the correlation expansion in $g_{\textrm{ph}}$ 
the equation of motion for the normal quasiparticle density $\langle \alpha^{\dag}_{{\bf k} } \alpha^{\phantom{\dag}}_{{\bf k}'} \rangle$, 
as obtained from the Heisenberg equation of motion, is given by
\begin{widetext}
\begin{eqnarray} \label{LongEq}
&&
i \hbar \frac{ d }{ dt } \left\langle \alpha^{\dag}_{\bf k } \alpha^{\phantom{\dag} }_{{\bf k}' } \right\rangle
=
\left(
R_{{\bf k}' } - R_{{\bf k}}
\right) \langle \alpha^{\dag}_{{\bf k}} \alpha^{\phantom{\dag}}_{{\bf k}' } \rangle
+
C_{{\bf k}' } \langle \alpha^{\dag}_{{\bf k}} \beta^{\dag}_{ {\bf k}' } \rangle
+
C^{\ast}_{{\bf k}} \langle \alpha^{\phantom{\dag}}_{{\bf k}' } \beta^{\phantom{\dag}}_{ {\bf k}} \rangle
\nonumber\\
&& \quad
-
\frac{ e \hbar}{2 m} 
\sum_{{\bf q}= \pm {\bf q}_{\textrm{p}} }
( 2 {\bf k} + {\bf q} ) \cdot {\bf A}_{{\bf q} } 
\Big[
 L^{+ }_{{\bf k}, {\bf q} } 
\langle \alpha^{\dag}_{{\bf k} + {\bf q} }\alpha^{\phantom{\dag}}_{{\bf k}' } \rangle
-
L^{+ }_{{\bf k}', - {\bf q} }
\langle \alpha^{\dag}_{{\bf k}} \alpha^{\phantom{\dag}}_{{\bf k}' - {\bf q} } \rangle
-
M^{- }_{{\bf k}, {\bf q} }
\langle \alpha^{\phantom{\dag}}_{{\bf k}' } \beta^{\phantom{\dag}}_{ {\bf k} + {\bf q} } \rangle
-
M^{- }_{{\bf k}' , - {\bf q} }
\langle \alpha^{\dag}_{{\bf k}} \beta^{\dag}_{ {\bf k}' -{\bf q} } \rangle
\Big]
\nonumber\\
&& \quad
- \frac{ e^2 }{2 m}
 \sum_{{\bf q} }
\left( \sum_{{\bf q}' = \pm {\bf q}_{\textrm{p}} } {\bf A}_{{\bf q} - {\bf q}' } \cdot {\bf A}_{{\bf q}' } \right)
\Big[
L^-_{{\bf k}, {\bf q} }
\langle \alpha^{\dag}_{{\bf k} + {\bf q} } \alpha^{\phantom{\dag}}_{{\bf k}' } \rangle
-
L^{-}_{{\bf k}' , - {\bf q} }
\langle \alpha^{\dag}_{{\bf k}} \alpha^{\phantom{\dag}}_{{\bf k}' - {\bf q} } \rangle
+
M^{+}_{{\bf k}, {\bf q} }
\langle \alpha^{\phantom{\dag}}_{{\bf k}' } \beta^{\phantom{\dag}}_{ {\bf k} + {\bf q} } \rangle
+
M^{+}_{{\bf k}' , - {\bf q} }
\langle \alpha^{\dag}_{{\bf k}} \beta^{\dag}_{ {\bf k}' - {\bf q} } \rangle
\Big] 
\nonumber\\
&& \quad
- 
g_{\textrm{ph} } 
\sum_{{\bf p} } 
D_{\bf p} 
\Big[
M^{+ }_{ {\bf k}' , - {\bf p} } \langle \alpha^{\dag}_{{\bf k}} \beta^{\dag}_{{\bf k}' - {\bf p}} \rangle 
+ M^{+}_{{\bf k}, {\bf p}} \langle \alpha^{\phantom{\dag}}_{{\bf k}' } \beta^{\phantom{\dag}}_{{\bf k} + {\bf p}} \rangle
+ L^{- }_{{\bf k}, {\bf p}} \langle \alpha^{\dag}_{{\bf k} + {\bf p}} \alpha^{\phantom{\dag}}_{{\bf k}' } \rangle
- L^{- }_{{\bf k}' , - {\bf p}} \langle \alpha^{\dag}_{{\bf k}} \alpha^{\phantom{\dag}}_{{\bf k}' - {\bf p}} \rangle
\Big] ,
\end{eqnarray}
\end{widetext}
where $R_{\bf k} = \varepsilon_{{\bf k}} ( 1 - 2 v^2_{\bf k} )+ 2 u_{\bf k} v_{\bf k} \textrm{Re} \Delta$, $C_{\bf k} = - 2 \varepsilon_{{\bf k}} u_{\bf k} v_{\bf k} + \Delta u^2_{\bf k} - \Delta^{\ast} v^2_{\bf k}$,  $L^{\pm}_{{\bf k}, {\bf k}'}=u_{\bf k} u_{{\bf k} + {\bf k}' } \pm v_{{ \bf k}} v_{{\bf k} + {\bf k}'}$, and $M^{\pm}_{{\bf k}, {\bf k}'}=v_{{\bf k}} u_{{\bf k} + {\bf k}' } \pm u_{{\bf k}} v_{{\bf k} + {\bf k}' }$. Comparing the first and the last line of Eq.~\eqref{LongEq}, one sees that
 the quasiparticle-phonon interaction at first order in the hierarchy simply leads to a nondiagonal energy renormalization.
The equations of motion for the remaining three quasiparticle densities, $\langle \beta^{\dag}_{ {\bf k} } \beta^{\phantom{\dag}}_{ {\bf k}^{\prime} } \rangle$, $\langle \alpha^{\dag}_{ {\bf k}} \beta^{\dag}_{ {\bf k}^{\prime}} \rangle$, and $\langle \alpha^{\phantom{\dag}}_{ {\bf k}} \beta^{\phantom{\dag}}_{ {\bf k}^{\prime}} \rangle$, which have a similar structure, are given in Appendix~\ref{appendixA}.

The time dependence of the coherent-phonon amplitude $D_{\bf p}(t)$ can be expressed in terms of a harmonic oscillator-type second-order 
differential equation (for details see Appendix~\ref{appendixA})
\begin{subequations} \label{harmOsc}
\begin{eqnarray} \label{harmOscA}
\left[ \frac{ d^2 }{d t^2} + \omega^2_{\textrm{ph}} \right] 
D_{\bf p}(t)
&=&
 \mathcal{F}_{\bf p} ( t) ,
\end{eqnarray}
with forcing term
\begin{eqnarray} \label{forcingTerm}
\mathcal{F}_{\bf p} (t)
&=&
- \frac{ 2 \omega_{\textrm{ph}} }{\hbar} g_{\textrm{ph}}
\sum_{{\bf k}} 
\Big[
M^{+}_{{\bf k}, {\bf p}}
\left( \langle \alpha^{\phantom{\dag}}_{{\bf k} + {\bf p} } \beta^{\phantom{\dag}}_{{\bf k}} \rangle - \langle \alpha^{\dag}_{{\bf k}} \beta^{\dag}_{{\bf k} + {\bf p}} \rangle \right)
\nonumber\\
&& \quad
+
L^{-}_{{\bf k}, {\bf p}} 
\left( \langle \alpha^{\dag}_{{\bf k}} \alpha^{\phantom{\dag}}_{{\bf k}+ {\bf p} } \rangle + \langle \beta^{\dag}_{{\bf k}+ {\bf p}} \beta^{\phantom{\dag}}_{{\bf k}} \rangle \right)
\Big] ,
\end{eqnarray}
\end{subequations}
which is purely real.
Within the framework of model~\protect{\eqref{model1}}, the equation of motion for the coherent-phonon amplitude $D_{\bf p}(t)$ is exact  up to higher-order corrections in the correlation expansion. It is worth noting, that at the next order in the hierarchy (i.e., at second order in $g_{\textrm{ph}}$) incoherent phonons and quasiparticle-phonon scattering processes are generated, 
which give rise to a finite lifetime of the coherent phonons  and which thereby lead to an
exponential damping of the coherent-phonon oscillations.
Focusing on time scales much shorter than the coherent phonon lifetime, we neglect in the following any finite lifetime effects due to quasiparticle-phonon or
phonon-phonon scattering processes.

Eq.~\eqref{LongEq} and the corresponding equations for the other three quasiparticle densities (see Appendix~\ref{appendixA}) together with Eq.~\eqref{harmOsc} form a closed set of coupled differential equations. In Sec.~\ref{sec:Simul} we solve numerically this set of equations  to determine the temporal evolution of the order parameter amplitude $| \Delta (t) |$ and the lattice displacement $U ( {\bf r}, t) $. 
Before doing so, we present in Sec.~\ref{sec:Coh} a qualitative analysis of the differential equation \eqref{harmOsc} and derive approximate solutions for different time scale regimes.

\section{Coherent phonon generation mechanism}\label{sec:Coh}

The equation of motion \eqref{harmOsc} for the coherent-phonon amplitude $D_{\bf p}(t)$ resembles the equation of a forced harmonic oscillator with driving force $\mathcal{F}_{\bf p}(t)$. The forcing term $\mathcal{F}_{\bf p}(t)$ is a function of the quasiparticle densities and implicitly depends on the optical excitation conditions, since both the normal and the anomalous quasiparticle densities are driven by the optical pump pulse. 
Hence, a rapid increase in the Bogoliubov quasiparticle distribution function due to optical excitation acts as a driving force for coherent-phonon oscillations. To make this more precise, let us express the general solution of the second order differential equation~\eqref{harmOsc} as 
\begin{eqnarray} \label{DpSol}
D_{\bf p} (t) = \int_{t_i}^t d t' \; \mathcal{F}_{\bf p} ( t' ) \frac{ \sin \left[ \omega_{\textrm{ph}} ( t - t' ) \right] }{ \omega_{\textrm{ph}} } ,
\end{eqnarray}
where we assumed the following initial conditions: $D_{\bf p} ( t_i) =0$ and $\frac{d}{d t} \, D_{\bf p} (t_i) = 0$, for all ${\bf p}$. Depending on the considered ordering of time scales, the 
time dependence of the driving force $\mathcal{F}_{\bf p}$ can be approximated by different functions.

\begin{figure}[tb]
\begin{center} 
\includegraphics[width=.5\textwidth,angle=-0]{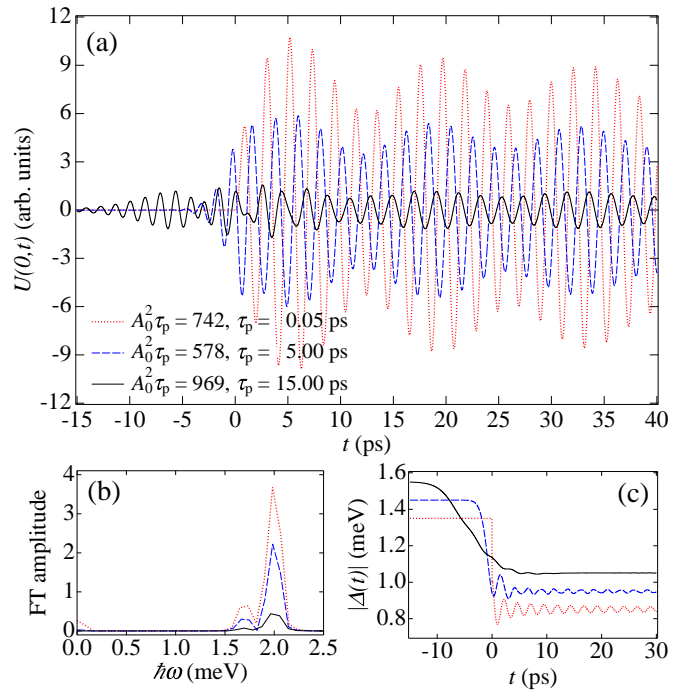}
\caption{
\label{fig2}
(Color online) Panel (a): Numerical simulation of the lattice displacement $U ( 0, t)$ versus time $t$ for three different pulse widths $\tau_{\textrm{p}}=15$ ps (solid black), $5$ ps (dashed blue), and $0.05$ ps (dotted red). The integrated pump pulse intensity for each trace is chosen such that $2 \Delta_{\infty} = 1.7$ meV. Here we take $\omega_{\textrm{ph}} = 2.0$ meV$/\hbar$. Panels (b) and (c) show the spectral distribution of $U (0, t)$ and the temporal evolution of $| \Delta (t) |$, respectively, for the same parameter values as in panel (a).
In panel (c), the curves are vertically shifted by multiples of $0.1$ meV.
}
\end{center}
\end{figure}

\begin{figure}[t!]
\begin{center} 
\includegraphics[width=.5\textwidth,angle=-0]{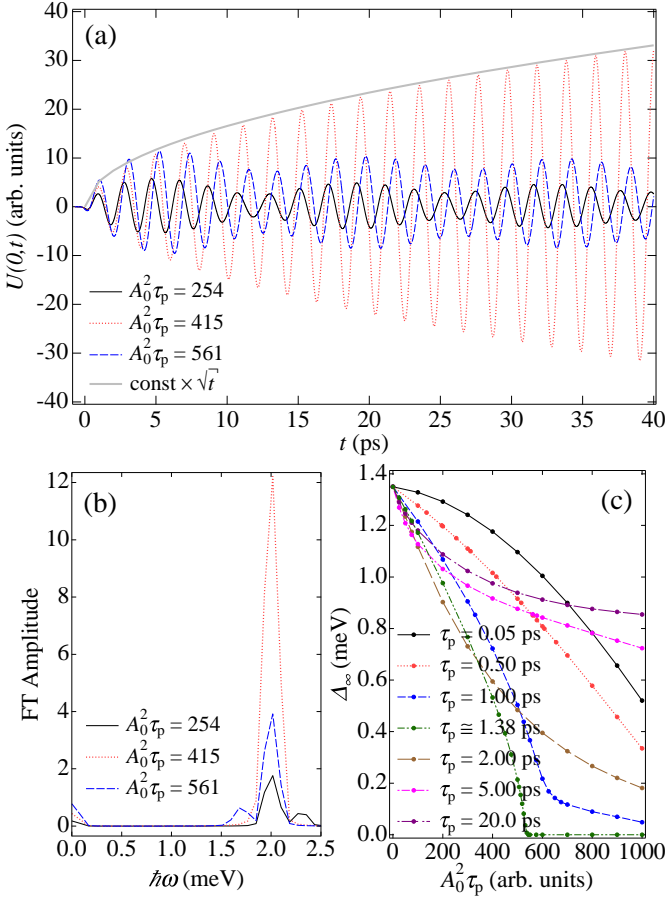}
\caption{
\label{fig3}
(Color online) Panel (a): 
Lattice displacement $U (0, t)$ as a function of time $t$ for resonant (dotted red: $A^2_{0} \tau_{\textrm{p}}  = 415$, $2 \Delta_{\infty} = 2.0$ meV) and off-resonant conditions 
(solid black: $A_0^2 \tau_{\textrm{p}} = 254$, $2 \Delta_{\infty} = 2.3$ meV;
dashed blue: $A_0^2 \tau_{\textrm{p}} =561$, $2 \Delta_{\infty} = 1.7$ meV).
Here we take $\tau_{\textrm{p}} = 0.5$ ps and $\omega_{\textrm{ph}} = 2.0$ meV$/\hbar$.
The gray curve shows the $\sqrt{t}$ dependence predicted by Eq.~\eqref{eqRes}.
Panel (b):
Spectral distribution of the coherent phonon oscillations for the same parameters as in panel (a).
Panel (c):
Asymptotic value of the order parameter $\Delta_{\infty}$ versus integrated pump pulse intensity
for seven different pulse widths $\tau_{\textrm{p}}$.
}
\end{center}
\end{figure}

First, we focus on the regime $\tau_{\textrm{p}} \ll \tau_{\Delta} \sim \tau_{\textrm{ph}}$, where both the quasiparticle and phononic subsystems evolve in a nonadiabatic manner and the phonon period $\tau_{\textrm{ph}}$ is of the same order of magnitude as the dynamical time scale of the order parameter $\tau_{\Delta}$. A number of recent publications have investigated this regime, albeit in the absence of phonon interactions.\cite{barankovPRL04,amin2004,Yuzbashyan05,YuzbashyanJPA05,Yuzbashyan06,papenkort07,papenkort08,papenkort09,papenkortConf09} Indeed, an exact solution has been derived for the dynamics of a BCS superconductor after an abrupt perturbation by, e.g., an interaction quench.\cite{barankovPRL04,amin2004,Yuzbashyan05,YuzbashyanJPA05,Yuzbashyan06} In particular, it has been shown that as $t \to \infty$, the absolute value of the order parameter $| \Delta (t) |$ approaches, in an oscillatory fashion, a constant value $\Delta_{\infty} < | \Delta (t_i) |$, i.e.
\begin{eqnarray} \label{oscEq}
| \Delta ( t ) |
=
\Delta_{\infty} 
+
\frac{ b }{\sqrt{t} } \cos \left( 2 \Delta_{\infty} t / \hbar + \phi \right) ,
\end{eqnarray}
where $b$ and $\phi$ are constants that depend on the initial state.\cite{Yuzbashyan06} The evolution of the normal and anomalous quasiparticle densities shows a similar oscillatory behavior with a $1/\sqrt{t}$ decay. As it turns out, the coupling to phonons does not qualitatively alter this time dependence, as long as the electron-phonon interaction strength is small compared to the superconducting gap amplitude. Hence, we approximate the forcing term in Eq.~\eqref{harmOsc} as
\begin{eqnarray} \label{FpAnsatz}
\mathcal{F}_{\bf p} (t) \simeq    \Theta (t) [ A_{\bf p} + B_{\bf p}   \cos \left( 2 \Delta_{\infty} t / \hbar \right) / \sqrt{t}  ],
\end{eqnarray}
with $\Theta( t)$ the Heaviside step-function.
Inserting Eq.~\eqref{FpAnsatz} into Eq.~\eqref{DpSol},
and assuming that the phonon frequency $\omega_{\textrm{ph}}$ is close to resonance with the order parameter oscillations, i.e., $\omega_{\textrm{d}} = | 2 \Delta_{\infty} / \hbar - \omega_{\textrm{ph}}| \ll \omega_{\textrm{ph}} $, we find that, to leading 
order in $\omega_{\textrm{d}} / \omega_{\textrm{ph}}$,
the coherent-phonon amplitude $D_{\bf p}(t)$ is given by
\begin{eqnarray} \label{eqBeat}
D_{\bf p}  (t)
\simeq
\frac{  B_{\bf p} }{ \omega_{\textrm{ph} } }
\sqrt{ \frac{ \pi }{ 2 \omega_{\textrm{d}} } } 
\Big[
\cos ( t \omega_{\textrm{ph}} ) S_2 ( t \omega_{\textrm{d}} ) + \sin ( t \omega_{\textrm{ph}} ) C_2 ( t \omega_{\textrm{d}} )  
\Big] ,
\hspace{-0.5cm}
\nonumber\\
\end{eqnarray}
for $t>0$, and where $S_2$ and $C_2$ denote the two Fresnel integrals.\cite{abramowitz} 
In other words, the time evolution of $D_{\bf p} (t)$ exhibits a beating-like phenomenon, i.e., $D_{\bf p} (t)$ oscillates with frequency $\omega_{\textrm{ph}}$ and an amplitude that is modulated by the Fresnel integrals (cf.\ Figs.~\ref{fig2} and~\ref{fig3}). Exactly at resonance, $\hbar \omega_{\textrm{ph}} = 2 \Delta_{\infty}$, the coherent-phonon amplitude takes the form
\begin{eqnarray} \label{eqRes}
D_{\bf p} (t)
&\simeq &
\frac{A_{\bf p}}{ \omega^2_{\textrm{ph}} } \left[ 1 - \cos ( \omega_{\textrm{ph}} t ) \right]
+ \frac{  B_{\bf p}}{ \omega_{\textrm{ph}} } 
\sqrt{t} \sin ( \omega_{\textrm{ph} } t )  
\nonumber\\
&&
+ 
 \frac{ B_{\bf p}}{ \omega_{\textrm{ph}} } 
 \int_0^t \frac{d t^{\prime}}{2 \sqrt{ t^{\prime} } } \sin \left[ \omega_{\textrm{ph}} ( t - 2 t^{\prime} ) \right]  ,
\end{eqnarray}
for $t>0$.
As $t$ increases, the second term quickly dominates in the above expression and, hence,  the amplitude of the oscillations  in $D_{\bf p}(t)$ grows like $\sqrt{t}$.
This is in excellent agreement with the numerical simulations presented in Sec.~\ref{sec:Simul} (cf. Fig.~\ref{fig3}).

Second, we consider the regime $  \tau_{\textrm{p}} ,\tau_{\Delta} \ll \tau_{\textrm{ph}}$, where the Bogoliubov quasiparticle oscillations average out on 
the time scale of the phonons. In this case, provided that $A_{\bf p}$ is not too small compared to $B_{\bf p}$ in Eq.~\eqref{FpAnsatz}, the forcing term $\mathcal{F}_{\bf p}(t)$ can be approximated by $\mathcal{F}_{\bf p} (t) \simeq A_{\bf p} \Theta (t)$.  Inserting this into Eq.~\eqref{DpSol} yields for the coherent-phonon amplitude $D_{\bf p}(t)$  
\begin{eqnarray} \label{eq_cos}
D_{\bf p} ( t) 
\simeq 
\frac{ A_{\bf p} }{ \omega_{\textrm ph}^2 } \left[ 1 - \cos ( \omega_{\textrm{ph}} t ) \right], \qquad \textrm{for $t>0$}.
\end{eqnarray}
Thus, the phonon oscillations are cosine-like, with the extrema lying at integer and half-integer multiples of the phonon period $\tau_{\textrm{ph}}$. 
The amplitude of the oscillations increases with decreasing phonon frequency $\omega_{\textrm{ph}}$.
Again, we find good agreement with the numerical results of Sec.~\ref{sec:Simul} (cf. Fig.~\ref{fig1}).
Note that when  $A_{\bf p} / B_{\bf p}$ becomes sufficiently small in Eq.~\eqref{FpAnsatz}, then  there appear
fast oscillations with frequency $2 \Delta_{\infty} / \hbar$  superimposed on the slow oscillations of Eq.~\eqref{eq_cos} 
(cf.\ solid black and dotted red curves in Fig.~\ref{fig1}).

\section{Numerical simulations}\label{sec:Simul}

In this section, we numerically solve the closed set of equations of motion, Eqs.~\eqref{LongEq}, \eqref{harmOsc},  \eqref{addEOF} and \eqref{EOFbeta}, 
both for the quasiparticle densities and the mean phonon amplitudes. From these quantities, the temporal evolution of the lattice displacement $U( {\bf r}, t)$ and of the order parameter amplitude $| \Delta (t) |$ is readily computed. Inspection of Eq.~\eqref{LongEq} shows that the largest entries in the quasiparticle density matrices are those confined to a band centered around the diagonal. That is, off-diagonal entries, such as e.g.\ $\langle \alpha^{\dag}_{\bf k } \alpha^{\phantom{\dag} }_{{\bf k} + n {\bf q}_{\textrm p} } \rangle$, are of order $| {\bf A}_0 |^n$. Hence, for sufficiently small $| {\bf A}_0 |$, the off-diagonal elements decrease rapidly as $n$ increases. To reduce the computational effort we therefore set all off-diagonal entries with $n>4$ to zero and, furthermore,  restrict ourselves to a one-dimensional wire geometry. It is important to note, however, that the phenomena discussed in this paper are qualitatively independent on the dimensionality of the system, as can be seen from the analytical analysis given in Sec.~\ref{sec:Coh}. In fact, for a related model it has been shown that \mbox{(quasi-)}one-dimensional simulations provide a good approximation for two- and three-dimensional superconductors.\cite{papenkort07}

For the numerical computations we use the following material parameters:\cite{footnoteParams}
superconducting gap in the initial state $\Delta (t_i) = 1.35$ meV, cut-off energy $\hbar \omega_c= 8.3$ meV, Fermi energy $E_{\textrm{F}} = 9479$ meV, effective electron mass $m= 1.9 m_0$, with $m_0$ the free electron mass. The optical pump pulse is centered at $t=0$ and has a central energy of $\hbar\omega_{\textrm{p}}= 3$ meV, which is of the same order, but slightly larger than $2 \Delta (t_i)$. As the initial state for the simulations we choose the equilibrium BCS ground state at zero temperature. 
We find that the lattice displacement $U(r,t)$ has  a quite weak dependence on position, exhibiting oscillations in time at all values of $r$. The frequency of these
oscillations does not depend on position, but their amplitude varies weakly with $r$.
Therefore, we choose to present in all the figures only the displacement for $r=0$. A detailed study of the r dependence of $U( r, t)$ will be presented elsewhere.

In the following, we adjust the pump pulse length $\tau_{\textrm{p}}$, the phonon frequency $\omega_{\textrm{ph}}$, 
as well as the integrated pump pulse intensity $A_{0}^2 \tau_{\textrm{p}} $, 
to explore different time scale regimes.

\paragraph{$\tau_{\Delta} \sim \tau_{\textrm{ph}}$:}  
We start with the most interesting case, namely the situation where the Bogoliubov quasipraticle oscillations are close to resonance with the phonon frequency, see Figs.~\ref{fig2} and \ref{fig3}.
In Figs.~\ref{fig2}(a) and \ref{fig2}(c), we plot $U (0, t)$ and $| \Delta (t) |$, respectively, for three different pump pulse lengths $\tau_{\textrm{p}}=0.05$, $5$, and $15$ ps.
The integrated pump pulse intensity for each curve in Fig.~\ref{fig2} is adjusted such that  $2\Delta_{\infty} = 1.7$ meV. This ensures that the order parameter oscillations are always close to resonance with the phonon frequency. 
When the quasiparticle subsystem is perturbed nonadiabatically ($\tau_{\textrm{p}} < \tau_{\Delta}$, i.e.\ $\tau_{\textrm{p}} = 0.05$ ps in Fig.~\ref{fig2} and  $\tau_{\textrm{p}} = 0.5$ ps in Fig.~\ref{fig3}),
the quasiparticle densities build up in a coherent manner while the system is out of equilibrium. This leads to rapid oscillations in the quasiparticle densities, and hence also in the order parameter, with an amplitude decaying as $1 / \sqrt{t}$ and a frequency that is determined by the asymptotic gap value $\Delta_{\infty}$ [see Fig.~\ref{fig2}(c) and Eq.~\eqref{oscEq}]. Consequently, the coherent phonons are driven by a sinusoidal forcing term. Thus, whenever the order parameter oscillations are close to resonance with the phonon mode (i.e., $| 2 \Delta_{\infty} / \hbar - \omega_{\textrm{ph}}| \ll \omega_{\textrm{ph}}$), we observe a pronounced beating phenomenon [cf.~Eq.~\eqref{eqBeat}].

Most importantly, we find that
the frequency of the order parameter oscillations can be tuned exactly to resonance by adjusting the integrated pump pulse intensity [cf.~Fig.~\ref{fig3}(c)]. 
This is demonstrated in Fig.~\ref{fig3}(a), which is the main result of our paper.
In this figure we plot the lattice displacement for $\hbar \omega_{\textrm{ph}} = 2$ meV and a fixed pulse duration $\tau_{\textrm{p}} = 0.5$~ps, but different integrated pulse intensities. At resonance, 
$A_0^2 \tau_{\textrm{p}}=415$ ($2 \Delta_{\infty} = 2$ meV), the amplitude of the phonon oscillations shows a square-root increase with $t$, which is in agreement with Eq.~\eqref{eqRes}. 
In Figs.~\ref{fig2}(b) and \ref{fig3}(b) we present the spectral distributions of the coherent phonon oscillations as obtained from the Fourier transforms of $U (0, t)$. The discussed behavior of $U (0, t)$ reflects itself in the Fourier transforms: 
for resonant condition we observe a strong single peak at $2$ meV, while for off-resonant condition there are two peaks, one at the phonon energy $\hbar \omega_{\textrm{ph}} = 2$ meV and the other at the frequency of the order parameter oscillations,  i.e.\ at  $2 \Delta_{\infty} = 1.7$ and $2.3$ meV, respectively.

\begin{figure}[tb]
\begin{center} 
\includegraphics[width=.49\textwidth,angle=-0]{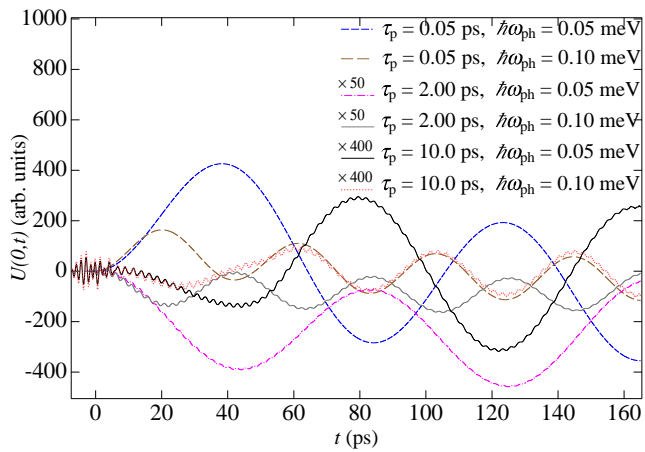}
\caption{
\label{fig1}
(Color online) 
Lattice displacement $U(0, t)$ as a function of time $t$ for 
three different pump pulse lengths $\tau_{\textrm{p}} = 0.05$, $2$, and $10$~ps
and two different phonon 
frequencies $\omega_{\textrm{ph}} = 0.1$ and $0.05$ meV$/\hbar$, 
corresponding to $\tau_{\textrm{ph}}  = 41$ and $83$ ps, respectively. 
The integrated pump pulse intensity for each curve is adjusted  such that $2 \Delta_{\infty} = 1.7$ meV.
The curves with $\tau_{\textrm{p}}=2$ and  $10$ ps have been multiplied by the constant factor $50$ and $400$, respecitvely.}
\end{center}
\end{figure}

In the regime, where the quasiparticles evolve adiabatically ($\tau_{\textrm{p}} > \tau_{\Delta}$, i.e.\ $\tau_{\textrm{p}} =5$ ps and $15$ ps in Fig.~\ref{fig2}), 
the pump pulse drives only the normal quasiparticle densities, $\langle \alpha^{\dag}_{\bf k } \alpha^{\phantom{\dag} }_{{\bf k}' } \rangle$ and $\langle \beta^{\dag}_{ {\bf k} } \beta^{\phantom{\dag}}_{ {\bf k}^{\prime} } \rangle$, but leaves the anomalous ones, $\langle \alpha^{\dag}_{ {\bf k}} \beta^{\dag}_{ {\bf k}^{\prime}} \rangle$ and $\langle \alpha^{\phantom{\dag}}_{ {\bf k}} \beta^{\phantom{\dag}}_{ {\bf k}^{\prime}} \rangle$, mostly unaffected.\cite{footnoteBasis} Thus,  the instantaneous value of the gap is almost fully determined at all times by the quasiparticle occupations, $\langle \alpha^{\dag}_{\bf k } \alpha^{\phantom{\dag} }_{{\bf k}' } \rangle$ and $\langle \beta^{\dag}_{ {\bf k} } \beta^{\phantom{\dag}}_{ {\bf k}^{\prime} } \rangle$, and the gap amplitude decreases monotonically from its initial equilibrium value $\Delta ( t_i ) $ to its final value $\Delta_{\infty}$ [black solid curve in Fig.~\ref{fig2}(c)].
As it turns out, in this situation the coherent phonons are still driven by a sinusoidal forcing term of the form~\eqref{FpAnsatz}, albeit with a much smaller amplitude. As a consequence, $U (r , t)$ still exhibits a beating phenomenon, but has a considerably smaller magnitude than in the nonadiabatic case.
Deep inside the adiabatic regime ($\tau_{\textrm{p}} \gg \tau_{\Delta}$) the coherent phonon oscillations eventually vanish completely.

In passing, let us also comment on the dependence of  $\Delta_{\infty}$ on the integrated pump pulse intensity $A_0^2 \tau_{\textrm{p}}$, which is shown 
in Fig.~\ref{fig3}(c) for seven different pulse widths $\tau_{\textrm{p}}$. The asymptotic gap value $\Delta_{\infty}$ is linear at small $A_0^2 \tau_{\textrm{p}}$ for all $\tau_{\textrm{p}}$, but deviates from this linear behavior at higher integrated intensities.
While the curves corresponding to short pump pulses ($\tau_{\textrm{p}} = 0.05$ and $0.5$ ps) exhibit a downward bend, 
those with longer pulse widths  ($\tau_{\textrm{p}} = 2.0$, $5.0$ and $20.0$ ps) flatten with increasing $A_0^2 \tau_{\textrm{p}}$.
(The curves with $\tau_{\textrm{p}}=1.0$ and $1.38$ ps  lie in between these two regimes, showing first a downward and then an upward bend.)
The downward bend is due to a quadratic term in the $A_0^2 \tau_{\textrm{p}}$ dependence resulting from two-photon processes. 
The flattening, on the other hand,  occurs because long pump pulses, with $\tau_{\textrm{p}} \gg 2 \pi / \omega_{\textrm{p}} \simeq 1.38$ ps, create sharp and narrow peaks in the quasiparticle distributions,
which, for sufficiently high intensities, leads to saturation due to Pauli blocking.\cite{papenkort07,papenkortConf09}
We observe that the integrated intensity above which Pauli blocking sets in decreases with increasing~$\tau_{\textrm{p}}$.

\begin{figure}[t!]
\begin{center} 
\includegraphics[width=.5\textwidth,angle=-0]{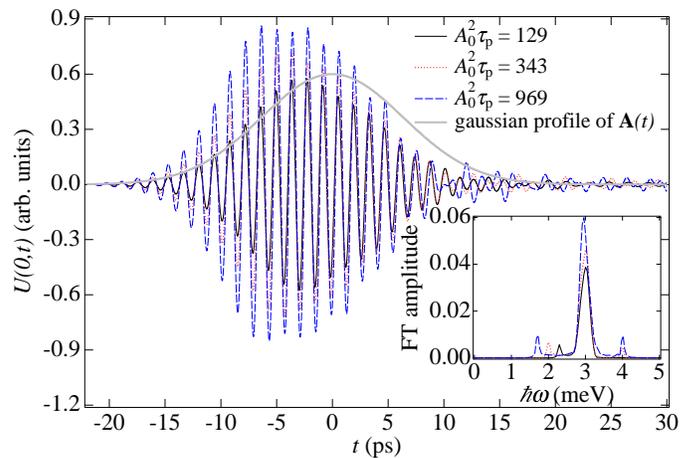}
\caption{
\label{fig4}
(Color online) Lattice displacement $U( 0, t)$ versus time $t$ for three different integrated pump pulse intensities $A_{0}^2 \tau_{\textrm{p}}= 129$, $343$, and $969$. 
These integrated intensities correspond to $2 \Delta_{\infty} = 1.7$, $2.0$, and $2.3$, respectively.
Here we choose $\tau_{\textrm{p}} = 15$ ps and $\omega_{\textrm{ph}} = 4$ meV$/\hbar$. The gray trace depicts the Gaussian time dependence of the pulse envelope. The inset shows the spectral distribution of the coherent phonon oscillations for the same parameters as in the main panel.}
\end{center}
\end{figure}

\paragraph{$\tau_{\Delta} , \tau_{\textrm{p}} \ll \tau_{\textrm{ph}}$:} 
In Fig.~\ref{fig1} we show  the lattice displacements $U ( 0, t)$ induced by  optical pump pulses with pulse lengths $\tau_{\textrm{p}} =0.05$, $2$ and $10$ ps for two different phonon energies  $\hbar\omega_{\textrm{ph}}=0.1$, and $0.05$ meV. This parameter choice corresponds to the case, where the rapid oscillations in the quasiparticle subsystem average out on the time scale of the phonons.  Since  $\tau_{\textrm{p}} \ll \tau_{\textrm{ph}}$, the phonons are perturbed by an almost instantaneous change in quasiparticle occupations, which leads to  cosine-like coherent phonon oscillations with frequency $\omega_{\textrm{ph}}$ and an amplitude that increases with decreasing phonon frequency $\omega_{\textrm{ph}}$ [cf.~Eq.~\eqref{eq_cos}]. As we go from the regime where the Bogoliubov quasiparticles are perturbed nonadiabaticallly ($\tau_{\textrm{p}} < \tau_{\Delta}$, i.e. $\tau_{\textrm{p}} = 0.05$ ps in Fig.~\ref{fig1}) to the regime where 
the quasiparticles are perturbed adiabatically ($\tau_{\textrm{p}} > \tau_{\Delta}$, i.e. $\tau_{\textrm{p}} = 10$ ps in Fig.~\ref{fig1}) the amplitude of the oscillations decreases quickly. For sufficiently long pump pulses,  eventually there appear fast oscillations with frequency $2 \Delta_{\infty} / \hbar$ on top of the slow oscillations with frequency $\omega_{\textrm{ph}}$
($\tau_{\textrm{p}} = 10$ ps in Fig.~\ref{fig1}).

\paragraph{$\tau_{\textrm{ph}} < \tau_{\Delta} \ll \tau_{\textrm{p}}$:} 
Lastly, we consider the case where both the quasiparticles and the phonons are perturbed in an (almost) adiabatic fashion and the coherent phonon oscillations are off resonance.
In Fig.~\ref{fig4}, the time dependence 
of $U (0, t)$ is shown for a pump pulse with length $\tau_{\textrm{p}} = 15$ ps 
and three different integrated pump pulse intensities.
The phonon energy is chosen to be $\hbar \omega_{\textrm{ph}} = 4$ meV, which is larger than $2 \left| \Delta (t_i) \right|$ and hence far away from resonance.
In spite of the almost adiabatic evolution of the system on the phonon time scale, coherent phonons are still being generated, albeit with a much smaller amplitude than in Figs.~\ref{fig2}-\ref{fig1}. Remarkably, the Fourier spectrum of the coherent phonon oscillations (inset in Fig.~\ref{fig4}) 
does not only show contributions at $2 \Delta_{\infty}$ and $\hbar \omega_{\textrm{ph}} = 4$ meV,
but also a third peak at $3$~meV, which is identical to the pump photon energy. 
The latter contribution is caused by large transient oscillations occurring in the time interval $\sim [ - \tau_{\textrm{p}}, + \tau_{\textrm{p}} ]$ during which the pump pulse acts on the system.
If the pump pulse frequency $\omega_{\textrm{p}}$ is chosen close to $\omega_{\textrm{ph}}$, the coherent phonon oscillations show some enhancement, 
i.e., the  coherent-phonon amplitude increases (almost) monotonically until $t \simeq 0$ ps, and then remains constant at its peak value even after
the pump pulse has passed  (not shown).

\section{Conclusions}

In this paper we have presented a theoretical investigation of the generation of coherent phonons in a superconductor by ultrafast laser pulses.
Using the density-matrix formalism, we have performed numerical simulations of the nonequilibrium dynamics of a BCS $s$-wave superconductor coupled to a single branch of optical phonons.
Based on both numerical and analytical arguments,
we have shown that sudden changes in the Bogoliubov quasiparticle densities created by the optical pump pulse act as a driving force for coherent phonon oscillations. For ultrafast laser excitations, the superconductor exhibits a nonadiabatic coherent dynamics that is characterized by rapid order parameter oscillations. We have found that the creation of coherent phonons is resonantly enhanced
when the period of these gap oscillations coincides with the phonon period.
In a pump probe experiment  this resonance condition can be achieved by tuning the frequency of the gap oscillations via a change in the integrated pump pulse intensity (see Fig.~\ref{fig3}).

The resonant coherent phonon generation mechanism discussed in this paper applies in principle to any BCS-type superconductor that has an optical phonon with phonon energy 
 of the same order as the superconducting gap. One interesting class of examples are superconductors which are close to a structural transition that is driven by a soft
optical phonon, i.e., e.g., CaC$_6$\cite{kimBoeriPRB06,emeryCaC6} or CaAlSi.\cite{imaiCaAlSi,boeriPRB08}  
The coherent phonon oscillations 
are experimentally observable, for example, as periodic modulations in time-resolved reflectivity measurements. Driving a superconductor into the regime of nonadiabatic coherent dynamics requires ultrashort laser pulses with frequencies of the order of the superconducting gap, i.e., in the terahertz regime. With the recent advent of ultrafast terahertz sources,\cite{williams07} we hope that it will be soon possible to perform time-resolved measurements on superconductors in the nonadiabatic regime and to test our theoretical predictions. The experimental observation of the discussed resonant coherent phonon oscillations would not only be interesting in itself, it could potentially also give useful information about the gap symmetry and the pairing mechanism of the superconductor.

\acknowledgements
This paper benefited from preliminary unpublished work done by M.~F\"orster. The authors thank I.\ Eremin, T.\ Papenkort, L.\ Boeri, and J.\ Bauer for discussions.
A.\ A.\ thanks the Max-Planck-Institut FKF Stuttgart for hospitality and financial support.

\appendix

\begin{widetext}

\section{Equations of motion}
\label{appendixA}

In this Appendix, we give the equations of motion for the quasiparticle density $\langle \alpha^{\dag}_{ {\bf k}} \beta^{\dag}_{ {\bf k}^{\prime}} \rangle$ and
$\langle \beta^{\dag}_{ {\bf k}} \beta^{\phantom{\dag}}_{ {\bf k}^{\prime}} \rangle$, 
and the mean phonon amplitude  $\langle b^{\phantom{\dag}}_{\bf p } \rangle$. In deriving these differential equations, we use the fact that
both $u_{{\bf k}}$ and $v_{{\bf k}}$ are real and time-independent, with $u_{{\bf k}} = \sqrt{ 1/2 ( 1+ \varepsilon_{\bf k} / E_{\bf k}) }$
and $v_{{\bf k}} = \sqrt{ 1/2 ( 1- \varepsilon_{\bf k} / E_{\bf k}) }$, and where $E_{\bf k} = \sqrt{ \varepsilon^2_{\bf k} +  | \Delta (t_i) |^2}$. 
As explained in the main text, we neglect terms of second- or higher-order in the correlation expansions in $g_{\textrm{ph}}$ and decouple 
phonon-assisted quantities according to, e.g., $\langle \alpha^{\dag}_{{\bf k} } \beta^{\dag }_{{\bf k}'} b^{\phantom{\dag}}_{{\bf p}} \rangle 
\simeq
\langle \alpha^{\dag}_{{\bf k} } \beta^{\dag}_{{\bf k}'} \rangle \langle b^{\phantom{\dag}}_{{\bf p}} \rangle$.
By use of Heisenberg's equation of motion, we find that the time dependence of  $\langle \alpha^{\dag}_{ {\bf k}} \beta^{\dag}_{ {\bf k}^{\prime}} \rangle$
is described by the following differential equation 
\begin{eqnarray}   \label{addEOF}
&&
i \hbar \frac{ d}{d t} \left\langle \alpha^{\dag}_{\bf k} \beta^{\dag}_{{\bf k}'} \right \rangle
=
- \left( R_{\bf k }+ R_{{\bf k}' } \right) \langle a^{\dag}_{{\bf k}} \beta^{\dag}_{{\bf k}' } \rangle
+  C^{\ast}_{{\bf k}' }  \langle \alpha^{\dag}_{{\bf k}} \alpha^{\ }_{{\bf k}' } \rangle
+ C^{\ast}_{{\bf k}} \left( \langle \beta^{\dag}_{{\bf k}'  } \beta^{\ }_{{\bf k}} \rangle - \delta_{{\bf k}'  ,  {\bf k}} \right) 
\nonumber\\
&& \;
- \frac{ e \hbar}{2 m}
\sum_{ {\bf q}  =  \pm {\bf q}_{\textrm{p}} }
( 2 {\bf k} + {\bf q} ) \cdot {\bf A}_{{\bf q} }  
\left[
 L^+_{{\bf k}, {\bf q}}
\langle  \alpha^{\dag}_{{\bf k} + {\bf q}} \beta^{\dag}_{{\bf k}'  } \rangle
-
L^+_{{\bf k}', - {\bf q} }
\langle \alpha^{\dag}_{{\bf k}} \beta^{\dag}_{  {\bf k}'  - {\bf q} } \rangle
+
M^-_{{\bf k}', - {\bf q} }
\langle \alpha^{\dag}_{{\bf k}} \alpha^{\ }_{{\bf k}'  - {\bf q} } \rangle
+
M^-_{{\bf k}, {\bf q} }
\langle \beta_{{\bf k}+{\bf q} }  \beta^{\dag}_{{\bf k}' }  \rangle 
\right]
\nonumber\\
&& \;
- \frac{e^2}{2m} \sum_{{\bf q} } \left( \sum_{{\bf q}'=\pm {\bf q}_{\textrm{p}} } {\bf A}_{{\bf q}-{\bf q}'  } \cdot {\bf A}_{{\bf q}'  }   \right)
\left[
L^-_{{\bf k}, {\bf q} }
\langle \alpha^{\dag}_{{\bf k}+{\bf q} } \beta^{\dag}_{{\bf k}'  } \rangle
+
L^-_{{\bf k}', - {\bf q} }
\langle \alpha^{\dag}_{{\bf k}} \beta^{\dag}_{ {\bf k}'  - {\bf q} } \rangle
+
M^+_{{\bf k}', - {\bf q} }
\langle \alpha^{\dag}_{{\bf k}} \alpha_{{\bf k}' -{\bf q} } \rangle
-
M^+_{{\bf k}, {\bf q} }
\langle  \beta^{\ }_{ {\bf k} + {\bf q} } \beta^{\dag}_{ {\bf k}' } \rangle
\right]
\nonumber\\
&& \;
-  
 g_{\textrm{ph}}
  \sum_{{\bf p} } 
D_{\bf p}
\Big[
M^{+}_{{\bf k}, {\bf p}} 
\left( \langle \beta^{\dag}_{{\bf k}' } \beta_{{\bf k} + {\bf p}}  \rangle   - \delta_{{\bf k}+{\bf p}, {\bf k}'}  \right)
+
L^{-}_{{\bf k}' , - {\bf p} } \langle  \alpha^{\dag}_{{\bf k}} \beta^{\dag}_{{\bf k}' - {\bf p}}  \rangle
+ 
L^{-}_{{\bf k}, {\bf p}} \langle \alpha^{\dag}_{{\bf k} + {\bf p}} \beta^{\dag}_{{\bf k}' }  \rangle 
+
M^{+}_{{\bf k}' ,  - {\bf p} } \langle \alpha^{\dag}_{{\bf k}} \alpha_{{\bf k}' - {\bf p}}   \rangle
\Big] ,
\end{eqnarray} 
where $R_{\bf k}$, $C_{\bf k}$, $L^{\pm}_{\bf k}$, and $M^{\pm}_{\bf k}$ are defined in Sec.~\ref{densityM}.
Note that the equation of motion for $\langle \alpha^{\phantom{\dag}}_{ {\bf k}} \beta^{\phantom{\dag}}_{ {\bf k}^{\prime}} \rangle$
can be obtained straightforwardly from Eq.~\eqref{addEOF} by complex conjugation.
The equation of motion for $\langle \beta^{\dag}_{ {\bf k}} \beta^{\phantom{\dag}}_{ {\bf k}^{\prime}} \rangle$ reads
\begin{eqnarray}\label{EOFbeta}
&& 
i \hbar \frac{d}{dt} \left\langle \beta^{\dag}_{{\bf k}} \beta^{\ }_{{\bf k}' } \right\rangle
=
\left( R_{{\bf k}' } - R_{{\bf k}} \right) \langle \beta^{\dag}_{ {\bf k}}  \beta_{ {\bf k}' }  \rangle
+ C_{{\bf k}'  } \langle \alpha^{\dag}_{{\bf k}' } \beta^{\dag}_{{\bf k}} \rangle
+ C^{\ast}_{{\bf k}} \langle \alpha_{{\bf k}} \beta_{ {\bf k}' }  \rangle
\nonumber\\
&& \;
 +  \frac{e \hbar}{2m}
\sum_{{\bf q} = \pm  {\bf q}_{\textrm{p}} } ( 2 {\bf k} - {\bf q}  ) \cdot {\bf A}_{{\bf q} }  
\left[
L^+_{{\bf k}, - {\bf q} }
\langle \beta^{\dag}_{ {\bf k} - {\bf q} } \beta_{ {\bf k}'  } \rangle
-
L^{+ }_{{\bf k}'  , {\bf q} }
\langle \beta^{\dag}_{ {\bf k}} \beta_{ {\bf k}' + {\bf q} } \rangle
- 
M^{-}_{{\bf k}, - {\bf q} } 
\langle \alpha_{{\bf k} - {\bf q} } \beta_{ {\bf k}' } \rangle
-
M^{- }_{{\bf k}' , {\bf q} }
\langle \alpha^{\dag}_{{\bf k}' + {\bf q} } \beta^{\dag}_{ {\bf k}} \rangle
\right]
\nonumber\\
&& \;
- \frac{ e^2 }{2 m} \sum_{{\bf q}  }
\left( \sum_{{\bf q}' = \pm {\bf q}_{\textrm{p}} } {\bf A}_{{\bf q} - {\bf q}' } \cdot  {\bf A}_{{\bf q}'}  \right)  
\left[
L^{-}_{{\bf k}, - {\bf q} }
\langle \beta^{\dag}_{ {\bf k} - {\bf q} } \beta^{\ }_{ {\bf k}'  } \rangle
-
L^{- }_{{\bf k}' , {\bf q} }
\langle \beta^{\dag}_{ {\bf k}}  \beta_{ {\bf k}' + {\bf q} } \rangle
+
M^{+ }_{{\bf k}' ,  {\bf q} }
\langle \alpha^{\dag}_{{\bf k}' + {\bf q} }  \beta^{\dag}_{ {\bf k}} \rangle
 +
  M^+_{{\bf k}, - {\bf q} } 
\langle \alpha_{{\bf k} - {\bf q} } \beta_{ {\bf k}' } \rangle
\right] 
\nonumber\\
&& \;
- g_{\textrm{ph}} \sum_{{\bf p}  }  
D_{\bf p}
\Big[
M^+_{{\bf k} , - {\bf p} } \langle \alpha^{\ }_{{\bf k} - {\bf p}} \beta^{\ }_{{\bf k}' } \rangle
+
 M^{+  }_{{\bf k}'  ,  {\bf p}} \langle \alpha^{\dag}_{{\bf k}' + {\bf p}} \beta^{\dag}_{{\bf k}}  \rangle
- 
 L^{-   }_{{\bf k}'  , {\bf p} } \langle \beta^{\dag}_{{\bf k}}  \beta_{{\bf k}' + {\bf p}} \rangle
+
 L^{-}_{ {\bf k}, -{\bf p} } \langle  \beta^{\dag}_{{\bf k} - {\bf p}} \beta^{\ }_{{\bf k}' }  \rangle
\Big] .
\end{eqnarray} 
We also present here the equation of motion for the mean phonon ampliuted $\langle b^{\phantom{\dag}}_{\bf p } \rangle$, which
is derived in a similar fashion as Eq.~\eqref{addEOF},
\begin{eqnarray} \label{meanPhonEOF}
&&
i \hbar \frac{ d}{d t} \langle b^{\phantom{\dag}}_{{\bf p}} \rangle
=
\hbar \omega_{\textrm{ph}} \langle b^{\phantom{\dag}}_{{\bf p}} \rangle
-  \frac{\hbar}{2 \omega_{\textrm{ph}} } \mathcal{F}_{\bf p}Ê( t) ,
\end{eqnarray}
where $ \mathcal{F}_{\bf p}Ê( t)$ is defined in Eq.~\eqref{forcingTerm}.
Again, we note that the equation of motion for  $\langle b^{\dag}_{-\bf p } \rangle$
is related to the one for $\langle b^{\phantom{\dag}}_{{\bf p}} \rangle$
by complex conjugation.
Adding the equations for $\langle b^{\phantom{\dag}}_{\bf p } \rangle$ and $\langle b^{\dag}_{-\bf p } \rangle$
and taking a time derivative, one  derives the equation of motion 
for the coherent-phonon amplitude $D_{\bf p}(t)$, Eq.~\eqref{harmOsc}.

As can be seen from Eqs.~\eqref{LongEq}, \eqref{addEOF}, and  \eqref{EOFbeta}, any element in the quasiparticle density matrices, such as e.g.\ $\langle \alpha^{\dag}_{\bf k } \alpha^{\phantom{\dag} }_{{\bf k} + n {\bf q}_{\textrm p} } \rangle$ is strongly coupled only to those elements with indices in the subspace $\left( {\bf k} + l {\bf q}_{\bf p} , \, {\bf k} + l {\bf q}_{\bf p}  +m {\bf q}_{\bf p}   \right)$, where $l$, $m$, and $n$ are integers. Elements with indices in different subspaces are only weakly coupled through the superconducting order parameter.
For the one-dimensional simulations we therefore discretized the momentum space by a one-dimensional grid with mesh size $| {\bf q}_{\textrm{p}} |$. Furthermore, we
approximate $( 2 {\bf k} + {\bf q} )  \cdot {\bf A}_{{\bf q}} $ in the second line of Eqs.~\eqref{LongEq}, \eqref{addEOF}, and  \eqref{EOFbeta} by
$2 {\bf k}_\textrm{F}  \cdot {\bf A}_{{\bf q}}  $, which is justified since the photon wave vector is much smaller than the Fermi momentum $\left| {\bf k}_{\textrm{F}} \right|$.
We use  a standard fourth-order Runge-Kutta technique to integrate the equations of motions given by Eqs.~\eqref{LongEq}, \eqref{harmOsc}, \eqref{addEOF} and \eqref{EOFbeta}. 

\phantom{a}
\end{widetext}


\begin{thebibliography}{99}
\bibitem{footnoteGph}
With this choice, the lattice displacement is proportional to the electron-phonon coupling constant, and hence $g_{\textrm{ph}}$
can be absorbed in the units of the lattice displacement.
\bibitem{footnoteParams}
We emphasize that the resonant coherent phonon generation mechanism discussed in this paper is independent of the particular choice of material parameters.
\bibitem{footnoteBasis}
Note that, throughout this paper, normal and anomalous quasiparticle densities are always defined with respect to the Bogoliubov-de Gennes basis in which the initial state is diagonal.
\end{thebibliography}
\end{document}